\documentclass[a4paper]{article}

\usepackage{color}
\usepackage{subfig}
\usepackage{url}

\usepackage{multirow}
\newcommand{\topcaption}{%
\setlength{\abovecaptionskip}{0pt}%
\setlength{\belowcaptionskip}{10pt}%
\caption}

\usepackage{INTERSPEECH2022}
\title{Glow-WaveGAN 2: High-quality Zero-shot Text-to-speech Synthesis and Any-to-any Voice Conversion}
\name{Yi Lei$^1$, Shan Yang$^2$, Jian Cong$^1$, Lei Xie$^{1*}$\thanks{* Lei Xie is the corresponding author. This work was supported by the National Key R \& D Program of China (2020AAA0108600).}, Dan Su$^2$}
\address{
  $^1$Audio, Speech and Language Processing Group (ASLP@NPU), School of Computer Science, \\ Northwestern Polytechnical University, Xi'an, China\\
  $^2$ Tencent AI Lab, China}
\email{ \{leiyi, lxie\}@nwpu.edu.cn, \{shaanyang, dansu\}@tencent.com, npujcong@mail.nwpu.edu.cn}

\begin{document}

\maketitle
\begin{abstract}
The zero-shot scenario for speech generation aims at synthesizing a novel unseen voice with only one utterance of the target speaker. Although the challenges of adapting new voices in zero-shot scenario exist in both stages -- acoustic modeling and vocoder, previous works usually consider the problem from only one stage. In this paper, we extend our previous Glow-WaveGAN to Glow-WaveGAN 2, aiming to solve the problem from both stages for high-quality zero-shot text-to-speech and any-to-any voice conversion. We first build a universal WaveGAN model for extracting latent distribution $p(z)$ of speech and reconstructing waveform from it. Then a flow-based acoustic model only needs to learn the same $p(z)$ from texts, which naturally avoids the mismatch between the acoustic model and the vocoder, resulting in high-quality generated speech without model fine-tuning. Based on a continuous speaker space and the reversible property of flows, the conditional distribution can be obtained for any speaker, and thus we can further conduct high-quality zero-shot speech generation for new speakers. We particularly investigate two methods to construct the speaker space, namely pre-trained speaker encoder and jointly-trained speaker encoder. The superiority of Glow-WaveGAN 2 has been proved through TTS and VC experiments conducted on LibriTTS corpus and VTCK corpus.

\end{abstract}
\noindent\textbf{Index Terms}: Zero-shot, speech synthesis, voice conversion, variational auto-encoder, flow model

\section{Introduction}
\label{sec:intro}

\begin{figure*}[!htb]
  \centering
  \vspace{-10pt}
  \includegraphics[width=\linewidth]{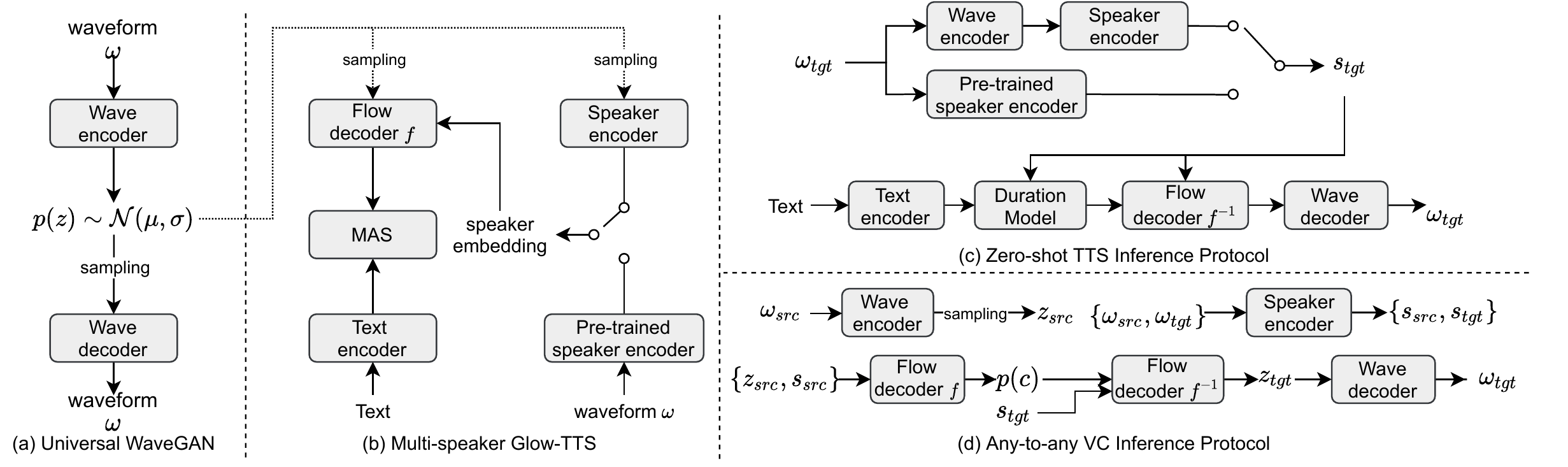}
  \caption{(a) Universal WaveGAN. (b) Multi-speaker acoustic model with optional speaker encoders. (c) Zero-shot TTS inference protocol from the waveform of target speaker $\omega_{tgt}$. (d) VC part which converts source speech $\omega_{src}$ with content $p(c)$ from speaker $s_{src}$ to target speech $\omega_{tgt}$ of speaker $s_{tgt}$.}
  \vspace{-25pt}
  \label{fig:system}
\end{figure*}

Recently, text-to-speech (TTS) and voice conversion (VC) have achieved significant improvements  with the rapid developments of sequence-to-sequence (seq2seq) based acoustic models~\cite{wang2017tacotron,arik2017deep,ren2019fastspeech, mayor2020fastvc} and high-quality neural vocoders~\cite{oord2016wavenet, prenger2019waveglow,kumar2019melgan,kong2020hifi}. With a large amount of studio-recorded high-quality data, it is easy to extend these models to multi-speaker scenarios~\cite{gibiansky2017deep,casanova2021sc}.
But since sizable training data of a new speaker is usually unavailable for customization, speaker adaptation in speech synthesis has attracted raising attention ~\cite{cooper2020zero,yan2021adaspeech,arik2018neural,luong2020nautilus}. Considering the amount of data for target speaker, adaptive speech synthesis can be divided into \textit{few-shot} and \textit{zero-shot} methods. Few-shot adaptation aims at generating new voices with a few samples of the target speaker, where fine-tuning the whole or a part of a pre-trained multi-speaker model is usually adopted~\cite{taigman2017voiceloop,luong2017adapting,chen2021adaspeech}. However, fine-tuning methods are time-consuming and will lead to individual speaker-dependent models, which are not friendly to speaker customization. Moreover, avoiding over-fitting is vital for adaptation with limited data. By contrast, zero-shot methods tend to generate a new voice with only one utterance of the specific speaker, without model fine-tuning in general. In this work, we focus on zero-shot speech generation, including text-to-speech (TTS) and voice conversion (VC).

For both zero-shot TTS and VC, there are two critical problems that affect the performance of speech generation: 1) with only one utterance available, the stable ability of acoustic models to produce natural speech with target speaker's identity and 2) the performance of the vocoder to reconstruct waveform of unseen speakers from predicted acoustic features. For the first problem, an effective way in zero-shot scenarios is to train a speaker-discriminative encoder to model the speaker space, where arbitrary voices can be generated through the constraint of speaker representation from the space with only one utterance of a novel speaker~\cite{nachmani2018fitting,pascual2019learning}.
In terms of the vocoder problem, recent high-quality neural vocoders are usually speaker-dependent, where the quality of the generated speech for unseen speakers will unavoidably deteriorate. Aiming to solve this problem, a WaveRNN-based universal vocoder~\cite{lorenzo2018towards} and a MelGAN-based universal vocoder~\cite{jang2020universal} was proposed, but there still exists quality gaps between seen and unseen speakers.

Recently, Glow-TTS~\cite{kim2020glow} proposed to model the distribution of the mel spectrogram with the flow to improve the quality of synthesized speech, where a pre-trained speaker encoder was further injected in SC-Glow-TTS~\cite{casanova2021sc} for zero-shot TTS. But fine-tuning the vocoder is still critical to the quality and similarity of unseen speakers. So the VITS~\cite{kim2021conditional} and our Glow-WaveGAN~\cite{cong2021glow} are proposed at the same time to avoid this problem by modeling a latent speech representation, but there exists many differences between them. In VITS, it proposes to reconstruct speech from the posterior distribution $p(z|c)$ of linear spectrogram through conditional VAE, where they utilize the reversible characteristics of flow to transform only the mean of text prior $c$ to improve the expressiveness of the prior distribution~\cite{kim2021conditional}. So they restrict the affine-coupling layers in flow to be volume-preserving~\cite{dinh2016density}, and it cannot produce speech of unseen speakers because of its discrete speaker space modeling. As for the two-stage Glow-WaveGAN, the goal of the WaveGAN is to learn a latent speech distribution $p(z)$ directly from waveform through an unconditional VAE and GAN, while the acoustic model aims at modeling the same distribution $p(z)$ with non-volume preserving flow from texts, which also avoid the mismatch problem~\cite{cong2021glow}. 

Since $p(z)$ in WaveGAN is unconditional, it could extract rich information of the waveform like timbre or contents, which means its potential ability on reconstructing waveform for unseen speakers. To this end, we propose Glow-WaveGAN 2 focusing on high-quality zero-shot text-to-speech synthesis and any-to-any voice conversion, where we build a universal WaveGAN and investigate different speaker encoders in flow on providing continuous speaker constraint $s$ on $p(z|s)$ for zero-shot scenarios. Given one utterance of an arbitrary speaker during inference, we can use the encoder of the universal WaveGAN to extract the speaker-related latent distribution $p(z)$, and then generate the speaker's voice with our conditional acoustic model and the decoder of WaveGAN. The experimental results on the VCTK and LibriTTS corpora show that the proposed methods can generate high-quality target voice without model fine-tuning in both zero-shot TTS and any-to-any VC.




\section{Method}
\label{sec:method}
As a substantial extension of our previous work~\cite{cong2021glow}, the proposed model mainly focuses on the zero-shot high-quality text-to-speech synthesis and voice conversion. As shown in Figure~\ref{fig:system}, 
the Glow-WaveGAN 2 contains three modules that will be elaborated in the following subsections: 1) a robust variational universal WaveGAN, which acts as both feature extractor and vocoder to extract the latent distribution $p(z)$ from speech and reconstruct speech from the sampled $z$ respectively; 2) a multi-speaker Glow-TTS~\cite{kim2020glow}, which generates latent representation $p(z|s)$ with the speaker identity $s$ as the conditioning constraint; and 3) an additional speaker encoder, which learns the speaker identity $s$ for adaptation.

\begin{table*}[ht]
\footnotesize
\centering
\topcaption{SECS of different models on different corpora, where GT-cross-spk is the SECS of different speakers. In VC scenarios, ``s2s'' indicates converting seen speakers to seen speakers, ``u2s'' means converting unseen speakers to seen speakers, ``s2u'' means converting seen speakers to unseen speakers, and ``u2u'' means converting unseen speakers to unseen speakers.}
\label{tab:secs}
 \resizebox{0.95\textwidth}{!}{
 \begin{tabular}{c|cc|cccc|cc|cccc}
 \toprule
Training data & \multicolumn{6}{c|}{LibriTTS}  & \multicolumn{6}{c}{VCTK} \\
\cline{1-1}\cline{2-7} \cline{8-13}
\multirow{2}{*}{Models} &             \multicolumn{2}{c|}{TTS} & \multicolumn{4}{c|}{VC} & \multicolumn{2}{c|}{TTS} & \multicolumn{4}{c}{VC} \\
                       \cline{2-3} \cline{4-7} \cline{8-9} \cline{10-13}
& seen & unseen & s2s & u2s & s2u & u2u & seen & unseen  & s2s & u2s & s2u  & u2u  \\ \midrule
GT-same-spk & \multicolumn{6}{c|}{0.830} & \multicolumn{6}{c}{0.815}  \\
GT-cross-spk & \multicolumn{6}{c|}{0.544}  & \multicolumn{6}{c}{0.551}  \\ \midrule
GlowTTS-HiFiGAN & 0.772 & - & 0.748 & - & - & - & 0.783 & - & 0.790 & - & -  \\
VITS & 0.807 & -  & 0.766 & - & - & - & 0.798 & - & \textbf{0.812} & - & -   \\\midrule
Glow-WaveGAN & 0.819 & - & 0.769 & -  & -  & - & \textbf{0.805} & - & 0.811 & - & -   \\
Glow-WaveGAN2-joint & 0.791 & 0.749 &  0.762 & 0.781 & 0.742 & 0.760  & 0.793 & 0.731 &  0.805 & 0.794 & 0.720 & 0.717  \\
Glow-WaveGAN2-pre & \textbf{0.822} & \textbf{0.784} & \textbf{0.771} & \textbf{0.802} & \textbf{0.771} & \textbf{0.807} &  0.804 & \textbf{0.774}  & 0.807 & \textbf{0.795} & \textbf{0.757} & \textbf{0.762} \\ \bottomrule
\end{tabular}
}
\vspace{-1.2em}
\end{table*}

\subsection{Universal WaveGAN}
\label{sec:ms-wavegan}
A robust vocoder for unseen speakers is critical for zero-shot speech generation. To achieve this goal, we build a variational auto-encoder (VAE)~\cite{Diederik2014auto} based universal WaveGAN model to conduct speech reconstruction, which has shown its strong ability to reconstruct unseen speakers in our previous work~\cite{cong2021glow}.

In the universal WaveGAN, the encoder aims at extracting the distribution of acoustic representation $z \sim p(z|w)$ from the waveform $w$, where $z$ also includes the speaker characteristics through a pitch predictor. Meanwhile, the decoder tends to reconstruct the speech $\hat{w}\sim p(w|z)$ from sampled $z$. To ensure the quality of reconstructed speech, we use adversarial training to model the $p(w)$ of the waveform. The training objective of the WaveGAN keeps the same as~\cite{cong2021glow}.

The encoder of WaveGAN can be treated as a robust feature extractor, and the extracted speech representation $z$ contains the speaker information. 
The decoder plays the role of vocoder to reconstruct high-quality speech waveform for both seen and unseen speakers. In this way, the proposed method guarantees the speech quality of zero-shot speech reconstruction.


\subsection{Multi-Speaker Glow-TTS}
\label{sec:ms-tts}
With the extracted latent speech representation, the skeleton of our acoustic model follows Glow-TTS~\cite{kim2020glow} to model the conditional distribution $p(z|t,s)$ from texts $t$, where $s$ is the conditional constraint to provide speaker identity. In detail, the text encoder transforms the text $t$ into a linguistic prior distribution $p(c|t)$. Considering the frame rate difference between the text and the latent $z$, monotonic alignment search (MAS) is adopted to align the prior $c$ and acoustic feature $z$. With the speaker embedding fed into the flow-based decoder $f$, the target $z$ from WaveGAN is transformed into the aligned prior distribution $p(c)$ through forward pass of $f$ during training.
Thus the log-likelihood of the target distribution $p(z)$ can be obtained through the flow-based decoder:
\begin{equation}
  \log P_Z(z|t,s) = logP_C(c|t) + \log \vert det \frac{\partial f^{-1}_{dec}(z,s)}{\partial_z } \vert
\end{equation}

Note that we specifically sample $z$ from the same $p(z)$ of the WaveGAN encoder in each training step, which mitigates the mismatch problem between the acoustic model and the vocoder to achieve high synthesis quality without fine-tuning. During inference, the predicted $\hat{z}$ is generated by the reverse decoding process $f^{-1}$ with the speaker condition from the text prior. Since the text encoder is speaker-independent, it's easy to conduct voice conversion through the forward $f$ and reverse $f^{-1}$ flow decoding with different speaker constraints~\cite{kim2020glow}, as shown in Figure~\ref{fig:system} (d).

\subsection{Speaker Representation}
\label{sec:spk-enc}
To achieve the zero-shot TTS and any-to-any VC, a key problem is how to model the speaker characteristics, especially for unseen speakers. In this paper, we investigate two alternative ways, a \textit{pre-trained encoder} and a \textit{jointly-trained encoder}, to build the speaker space in the acoustic model to produce speaker-dependent $z$ for unseen speakers.

A straightforward way to represent speakers is to utilize a pre-trained speaker encoder based on the speaker embedding extraction in speaker verification~\cite{cooper2020zero,jia2018transfer,snyder2018x,wan2018generalized}, through which we can obtain the speaker representation $s$ by only one utterance of a few seconds to generate speech of unseen speakers. Besides, a jointly-trained encoder based on speaker classification is also adopted in our work to learn $s$ from $p(z)$ within the Glow-TTS model. With the jointly-trained speaker module, we optimize the acoustic model with an extra cross entropy objective. Specifically, 
we sample two vectors $\{z_1, z_2\}$ from the same distribution $p(z)$, where $z_1$ and $z_2$ are used to train the acoustic model and speaker classification module respectively.

\subsection{Zero-shot TTS and VC}
\label{sec:zeroshot}
As shown in Figure~\ref{fig:system} (c) and (d), given an utterance of any target speaker during inference, the speaker identity $s_{tgt}$ can be extracted from the speaker encoder, which can be treated as the speaker constraint in our model. At inference time of zero-shot TTS, $s_{tgt}$ is utilized as the condition for the acoustic model to generate the sampled $z$ containing the target speaker information. At inference time of zero-shot VC, the source speaker $s_{src}$ is conditioned for the flow module to generate the speaker-independent linguistic prior distribution$p(c)$, which eliminates the source speaker information. Then $p(c)$ is fed into the reversed flow conditioned on the target speaker $s_{tgt}$ to generate the target sampled $z_{tgt}$. In this way, zero-shot TTS and VC can be achieved for generating speech of any speaker.

\vspace{-5pt}

\section{Experiments and results}
\label{sec:exp}

\subsection{Basic Setup}
\vspace{-3pt}
We conduct experiments on two different datasets named VCTK~\cite{veaux2016superseded} and LibriTTS~\cite{zen2019libritts} at 24 kHz. The VCTK corpus contains 109 English speakers, where we reserved 3 male and 3 female speakers as unseen speakers. As for the LibriTTS corpus, we use 1,151 speakers from the \textit{train-clean-360} and \textit{train-clean-100} subsets to train different systems, while the 39 speakers from the \textit{test-clean} subset are treated as unseen speakers. The jointly-trained speaker encoder consists of two 1d-convolution layers with kernel size 5, followed by the layer normalization and dropout. Finally, another convolution layer with 128 channels is adopted to extract the speaker embedding through mean pooling. As for the pre-trained speaker encoder, we utilize the voice encoder in the Resemblyzer tool~\cite{jia2018transfer,wan2018generalized} to extract a 256-dimensional speaker representation.

To evaluate the capability of the proposed Glow-WaveGAN 2 in zero-shot speech generation~\footnote{Audio samples can be found at \url{https://leiyi420.github.io/glow-wavegan2/}}, we set up two state-of-the-art models as baselines: (1) \textbf{GlowTTS-HiFiGAN}, which contains a multi-speaker flow-based acoustic model and the GAN-based vocoder to reconstruct mel-spectrogram; and (2) \textbf{VITS}~\cite{kim2021conditional}, which conducts multi-speaker synthesis in an end-to-end manner. Both of them utilize speaker ID as the speaker condition of the acoustic model. As for the Glow-WaveGAN family, we build different models for evaluation: (1) the basic\textbf{ Glow-WaveGAN} with explicit speaker labels like the above baselines, (2) the \textbf{Glow-WaveGAN2-joint} with jointly-trained speaker encoder, and (3) the \textbf{Glow-WaveGAN2-pre} with pre-trained speaker encoder.

\begin{table*}[ht]
\centering
\topcaption{MOS scores of different systems on two corpora for TTS and VC with 95\% confidence interval.}
\label{tab:mos}
 \resizebox{1\textwidth}{!}{
 \begin{tabular}{c|cc|cccc|cc|cccc}
\toprule
Training data & \multicolumn{6}{c|}{LibriTTS}  & \multicolumn{6}{c}{VCTK} \\
                       \cline{1-1} \cline{2-7} \cline{8-13}
\multirow{2}{*}{Models}   & \multicolumn{2}{c|}{TTS} & \multicolumn{4}{c|}{VC} & \multicolumn{2}{c|}{TTS} & \multicolumn{4}{c}{VC} \\
\cline{2-3} \cline{4-7} \cline{8-9} \cline{10-13}
& seen & unseen   & s2s & u2s & s2u & u2u & seen & unseen  & s2s & u2s & s2u  & u2u \\ \midrule
Grount-truth & \multicolumn{6}{c|}{4.38 $\pm$ 0.08} & \multicolumn{6}{c}{4.45 $\pm$ 0.07}  \\\midrule
GlowTTS-HiFiGAN & 3.36 $\pm$ 0.13 & -  & 3.45 $\pm$ 0.12 & - & - & - & 3.44 $\pm$ 0.11 & - & 3.55 $\pm$ 0.09 & - & - & - \\
VITS & 3.77 $\pm$ 0.11 & -  & 3.68 $\pm$ 0.13 & - & - & - & 3.83 $\pm$ 0.08 & - & 3.77 $\pm$ 0.09 & - & -  & - \\\midrule
Glow-WaveGAN & \textbf{3.78 $\pm$ 0.10} & - & 3.63 $\pm$ 0.11 & -  & -  & - & 3.85 $\pm$ 0.09 & - & 3.80 $\pm$ 0.08 & - & - & - \\
Glow-WaveGAN2-joint & 3.73 $\pm$ 0.12 & \textbf{3.69 $\pm$ 0.14}  & \textbf{3.78 $\pm$ 0.10} & 3.59 $\pm$ 0.14 & \textbf{3.53 $\pm$ 0.12} & 3.65 $\pm$ 0.15 &  \textbf{3.86 $\pm$ 0.11} & \textbf{3.87 $\pm$ 0.12} & \textbf{3.82 $\pm$ 0.08} & \textbf{3.73 $\pm$ 0.09} & \textbf{3.64 $\pm$ 0.11} & \textbf{3.75 $\pm$ 0.12} \\
Glow-WaveGAN2-pre & 3.69 $\pm$ 0.09 & 3.65 $\pm$ 0.11 & 3.73 $\pm$ 0.08 & \textbf{3.62 $\pm$ 0.09} & 3.51 $\pm$ 0.12 & \textbf{3.68 $\pm$ 0.11} & 3.81 $\pm$ 0.14 & 3.78 $\pm$ 0.09 & 3.78 $\pm$ 0.08 & 3.66 $\pm$ 0.11 & 3.58 $\pm$ 0.12 & 3.69 $\pm$ 0.14  \\ \bottomrule
\end{tabular}
}
\vspace{-1.2em}
\end{table*}

\subsection{Speaker Similarity Evaluation}
\label{sec:obj}
\vspace{-3pt}
We first evaluate the speaker similarity of the synthesized speech for TTS and VC between different models in the objective manner, which is critical to the performance of zero-shot speech generation. Following~\cite{casanova2021sc,choi2020attentron}, we treat the Speaker Encoder Cosine Similarity (SECS) between the speaker embeddings extracted from synthesized and ground-truth audios as an objective measure, where SECS scores range from $0$ to $1$ and a higher score means higher speaker similarity.

Table~\ref{tab:secs} shows the SECS results of different models. Note that the two baselines and the Glow-WaveGAN system adopt the explicit speaker labels to model the speaker identity, so they can only conduct speech synthesis and voice conversion for seen speakers. For the results of both TTS and VC on seen speakers, we find the VITS and the Glow-WaveGAN models achieve higher similarity than the GlowTTS-HiFiGAN model, especially in the LibriTTS dataset. And the Glow-WaveGAN family is slightly better than VITS for seen speakers in general.

For the zero-shot TTS of proposed models, the results show that there exists a gap in speaker similarity between seen and unseen speakers for the Glow-WaveGAN2-joint model. We believe that this is mainly because the jointly-trained speaker module only learns a weak speaker space on the limited training speakers since we can find such a gap is smaller in LibriTTS than that of VCTK, where LibriTTS has more speakers. Based on this assumption, we also evaluate the zero-shot TTS of the system Glow-WaveGAN-pre with the pre-trained speaker encoder, where the gap between seen and unseen speakers becomes obviously alleviated.

As for the any-to-any voice conversion task, we find that Glow-WaveGAN2-pre outperforms Glow-WaveGAN2-joint in general. The results of LibriTTS show that the Glow-WaveGAN2-pre model achieves similar SECS scores on the unseen target speakers to the seen targets. And it is worth noticing that the SECS score declines when the source speaker is seen, from which we argue that the model may still maintain the identity of source training speakers in the VC procedure. While for the VCTK corpus, the proposed methods achieve similar scores for the same target speaker, whether the source speaker is seen or unseen.

\begin{figure}[ht]
        \centering
        \includegraphics[width=\linewidth]{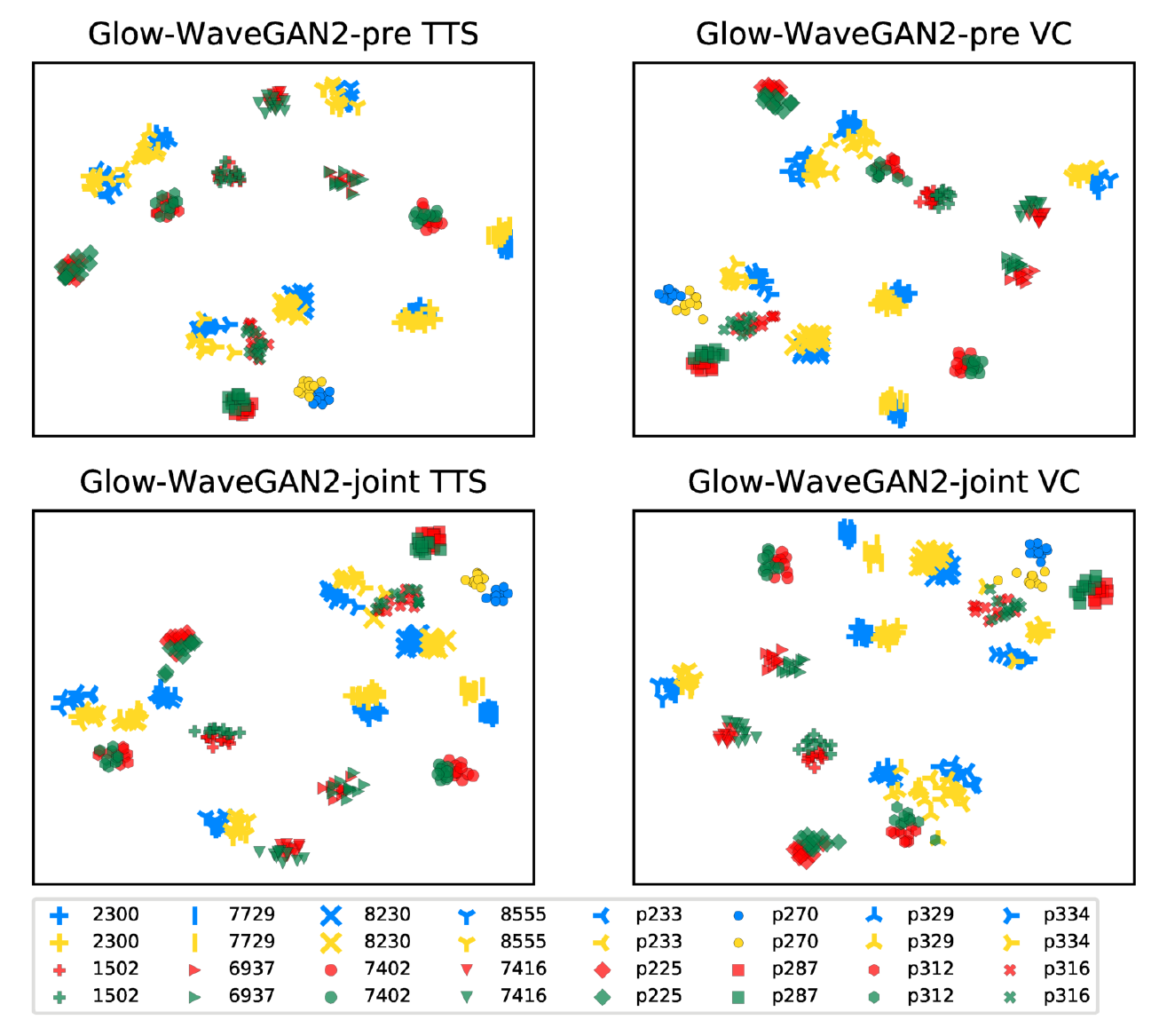}
        \caption{Speaker visualization of ground-truth and generated speech, where different shapes represent different speakers. The blue and yellow dots represent the \textit{unseen} ground-truth and generated speakers, and the red and green dots indicate \textit{seen} ground-truth and generated speakers.}
         \vspace{-20pt}
        \label{fig:tsne}
\end{figure}

\subsection{Speaker Representation Visualization}
\label{sec:visual}
\vspace{-3pt}

To further investigate the speaker similarity of the synthetic speech by the proposed methods individually trained with different corpora, the speaker embeddings of generated speech are visualized through t-SNE~\cite{van2008visualizing} as shown in Figure~\ref{fig:tsne}. For the seen (green and red dots) and unseen (blue and yellow dots) speakers on both TTS and VC tasks, the generated audios and corresponding reference speech of each speaker can form a distinct cluster in our two proposed methods, which shows that the proposed methods can effectively model and control the speaker identities. 

Compared with the indivisible clusters of unseen speakers formed from ground-truth and synthesized speech in the Glow-WaveGAN2-pre model with the pre-trained speaker encoder, there still exists boundaries of generated and ground-truth audios for some speakers in the Glow-WaveGAN2-joint model, especially for the smaller training corpus, i.e. VCTK, where the speaker's name begins with ``p'' in Figure~\ref{fig:tsne}.

\subsection{Speech Naturalness Evaluation}
\vspace{-3pt}

To evaluate speech naturalness and quality of different systems, we conduct Mean Opinion Score (MOS) tests, as shown in Table~\ref{tab:mos}. In each MOS test, there are 20 listeners rating 20 randomly chosen utterances for each model. In general, the MOS scores of VCTK are better than that of LibriTTS due to its better recording quality.

The MOS results of both TTS and VC on seen speakers demonstrate that the Glow-WaveGAN family and the VITS model have  obviously higher scores than the GlowTTS-HiFiGAN model, which comes from the mismatch problem mentioned in Section~\ref{sec:intro}. As for the unseen speakers in TTS, both the two proposed models can achieve similar MOS scores to that of the seen speakers, which indicates the effectiveness of the proposed models for zero-shot TTS. In the VC scenarios, the MOS results indicate that there is no significant difference between the Glow-WaveGAN family and the VITS model on seen speakers, where they both achieve satisfactory MOS scores on the VC task. When the source or the target speakers are unseen, the MOS results show that the proposed Glow-WaveGAN 2 models can also produce high-quality converted speech with only one utterance of the target speaker. Therefore, the experimental results demonstrate that our proposed models can generate high-quality speech in both zero-shot TTS and VC.

\subsection{Cross-dataset Evaluation}
\vspace{-3pt}

\begin{table}[]
\small
\centering
\topcaption{SCES and MOS results with 95\% confidence interval for cross-dataset evaluation. }
\label{tab:cross}
 \resizebox{0.49\textwidth}{!}{
\begin{tabular}{c|c|c|cc|cc}
\toprule
\multirow{2}{*}{Training data}   & \multirow{2}{*}{Testing data}  & \multirow{2}{*}{model}  & \multicolumn{2}{c|}{SECS}  & \multicolumn{2}{c}{MOS} \\ 
\cline{4-7} 
&&&TTS &VC &TTS &VC \\ \midrule
\multirow{2}{*}{LibriTTS}
    & \multirow{2}{*}{VCTK}
        & joint & 0.728 & 0.693  & \textbf{3.59 $\pm$ 0.11} & 3.54 $\pm$ 0.13 \\
    &   & pre  & \textbf{0.812} & \textbf{0.723} & 3.48 $\pm$ 0.12 & \textbf{3.57 $\pm$ 0.10} \\ \midrule
\multirow{2}{*}{VCTK}
    & \multirow{2}{*}{LibriTTS}
        & joint &  0.650 & 0.635 & \textbf{3.63 $\pm$ 0.13}  &  \textbf{3.31 $\pm$ 0.09} \\
    &   & pre   & \textbf{0.731} & \textbf{0.661} & 3.56 $\pm$ 0.08  &  3.25 $\pm$ 0.14 \\  \bottomrule
\end{tabular}
}
\vspace{-1.8em}
\end{table}

To further evaluate the generalization ability of our proposed models in zero-shot speech generation, we also calculate the SECS scores and conduct MOS tests across datasets. Specifically, the target speaker is from another dataset different from model training in TTS or the target and source speakers are both from another dataset different from model training in VC. Results are summarized in Table~\ref{tab:cross}. For speaker similarity, we find the model trained on the LibriTTS corpus has high SECS scores for cross-dataset evaluation since it can learn richer speaker space with more speakers. As for the subjective evaluation, it is interesting that we find the quality of generated speech from Glow-WaveGAN2-joint is better than that from Glow-WaveGAN2-pre in cross-dataset evaluation. We conjecture the reason is that the speaker constraint $s$ and the target $z$ are directly from the $p(z)$ of speech, which means that they may contain the channel information of the speech that affects the quality of generated speech in both training and inference.

\section{Conclusions}
In this paper, we propose Glow-WaveGAN 2, aiming at generating high-quality speech for zero-shot speech synthesis and any-to-any voice conversion. Specifically, we utilize a universal WaveGAN to build a robust feature extractor and high-quality universal vocoder. And the goal of flow-based multi-speaker acoustic model is to model the latent distributions conditioned on speaker constraints. We explore different speaker modeling strategies, and the results show that the proposed methods can produce high-quality speech in terms of naturalness and similarity for zero-shot speech generation.

\bibliographystyle{IEEEtran}

\bibliography{main}

\end{document}